# Basis Function feedforward for Position-Dependent Systems

M.J. van Haren[1], L.L.G. Blanken[1,2] and T.A.E. Oomen[1]

[1]Eindhoven University of Technology, the Netherlands

[2]Sioux Technologies, Eindhoven, the Netherlands

m.j.v.haren@tue.nl

**Abstract**

Feedforward for motion systems is getting increasingly more important to achieve performance requirements. This leads to a situation where position-dependent effects cannot be neglected anymore, where some examples can be seen in Figure 1.

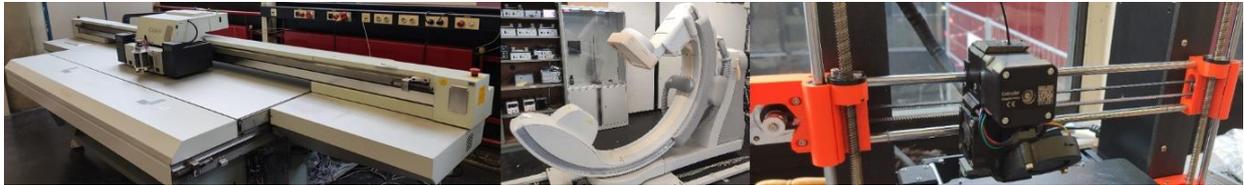

*Figure 1: Example position-dependent systems. Left: position-dependent flatbed printer. Center: position-dependent image guided therapy system. Right: timing-belt pulley system with position-dependent stiffness.*

Feedforward for motion systems typically consists of basis functions and parameters, e.g., mass and snap feedforward,

$$u_{ff} = m\frac{d^2}{dt^2}r + \delta\frac{d^4}{dt^4}r.$$

However, for position-dependent systems, feedforward should include position-dependent terms.

**Approach:** Learn position-dependent feedforward parameters in a different domain to achieve a less complex parameterization, i.e.,

$$u_{ff} = \frac{d^2}{dt^2}\left(\underbrace{m(\rho)}_{\theta_1(\rho)}\underbrace{1}_{\psi_1}r + \underbrace{\delta(\rho)}_{\theta_2(\rho)}\underbrace{\frac{d^2}{dt^2}r}_{\psi_2}\right).$$

The parameters are learned by applying recent advances in machine learning techniques, more specifically, kernel regularized system identification as seen in Pilonetto *et al.* [1] and Blanken *et al.* [2]. The kernel used specifies a prior on the feedforward parameter, for example smoothness or periodicity, as a function of position.

**Example:** Consider the timing-belt pulley system with position-dependent stiffness in Figure 2. This system represents the timing-belt pulley system of the 3D printer seen in the right of Figure 1.

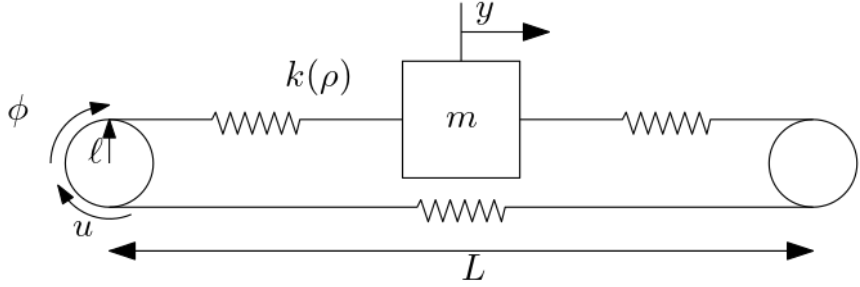

*Figure 2: Timing-belt pulley system considered, with position-dependent spring stiffness $k(\rho)$.*

The LPV input-output dynamics of this system, assuming one lumped spring with stiffness $k(\rho)$, are described by

$$\left(\frac{mJ}{r}\frac{d^2}{dt^2} + d\left(\frac{J}{r} + mr\right)\frac{d}{dt} + k(\rho)\left(\frac{J}{r} + mr\right)\right)y = \left(d\frac{d}{dt} + k(\rho)\right)\iint u\ dt^2.$$

Note that differentiation on both sides of this equation would lead to a complex expression due to the chain rule of differentiation, i.e., the operator $k(\rho)$ does not commute. The parameters are learned for the timing-belt pulley system with

$$k(\rho) = \frac{2L \cdot E \cdot A}{\rho\ell(2L - \rho\ell)} + 100 \cdot \sin(5 \cdot \rho),$$

where $E$ is the elasticity modulus and $A$ is the cross-sectional area of the timing belt. The developed approach that estimates the position-dependent snap feedforward parameter can be seen in Figure 3.

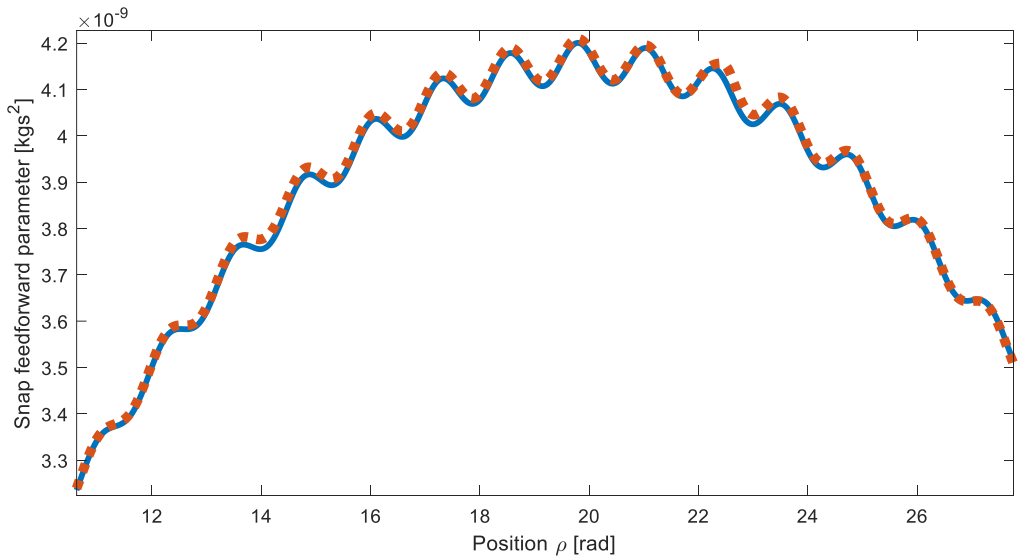

*Figure 3: Learned (dotted orange) and true (solid blue) position-dependent snap feedforward parameter of the pulley system.*

In terms of tracking performance, the normalized maximum error and error 2-norms are seen in Table 1.

Table 1. Normalized maximum tracking error and error 2-norm for a point-to-point tracking motion for the conventional position-independent feedforward and the developed approach.

| Method | Maximum Error [-] | Error 2-norm [-] |
| --- | --- | --- |
| Position-independent feedforward | 1 | 1 |
| Developed approach | 0.024 | 0.027 |

**References**

[1] Pillonetto G, Dinuzz, F, Chen T, de Nicolao G and Ljung L 2014. Kernel methods in system identification, machine learning and function estimation: A survey *Automatica.* **50** 657-682

[2] Blanken L and Oomen T 2020. Kernel-based identification of non-causal systems with application to inverse model control *Automatica.* **114** 108830